
\normalbaselineskip=20pt 
\baselineskip=20pt 
\magnification=1200 
\hsize 15.3true cm
\vsize 21.0true cm

\def\lsim{\mathrel{\rlap{\lower4pt\hbox{\hskip1pt$\sim$}}
    \raise1pt\hbox{$<$}}}         
\def\gsim{\mathrel{\rlap{\lower4pt\hbox{\hskip1pt$\sim$}}
    \raise1pt\hbox{$>$}}}         

\def\pom{\hbox{I$\!$P}}


\def\NP#1#2#3{ Nucl. Phys. {\bf #1} (#2) #3}

\def\PL#1#2#3{ Phys. Lett. {\bf #1} (#2) #3}

\def\PR#1#2#3{ Phys. Rev. {\bf #1} (#2) #3}
\def\PRL#1#2#3{ Phys. Rev. Lett. {\bf #1} (#2) #3}

\def\ZP#1#2#3{ Z. Phys. {\bf #1} (#2) #3}

\def\JETP#1#2#3{ Sov. Phys. JETP {\bf #1} (#2) #3}

\def\NC#1#2#3{ Nuovo Cimento {\bf #1} (#2) #3}


\rightline{ DFTT 72/95}
\centerline{\bf A simple recipe to detect possible C-Odd effects in high
energy $\bar p p$ and $pp$ .}

\vskip 24pt

\centerline{M. Giffon}
\centerline{Institut de Physique Nucl\'eaire de Lyon }
\centerline{Universit\'e Claude Bernard Lyon - FRANCE}

\centerline{E. Predazzi}

\centerline{ Dipartimento di Fisica Teorica - Univ di Torino. Torino - ITALY}

\centerline{and Sezione di Torino dell'Istituto Nazionale di Fisica Nucleare}
\medskip

\centerline{and}
\centerline{A. Samokhin}
\centerline{Institute for High Energy Physics. Protvino - RUSSIA}

\vskip 24pt

\noindent{{\bf Summary.} We provide a theorem to suggest that $t=0$ data
may already be sufficient to detect possible asymptotic C-odd ({\it Odderon})
contributions. This can be done
by comparing $\bar p p$ and $pp$ $t=0$ observables such as total
cross sections, forward angular distributions and ratios of
real to imaginary forward amplitudes for which well defined model independent
correlations {\it must} exist which could already show up at RHIC energy but
definitely at LHC energies.}

\vfill \eject

A long debated and still unresolved puzzle of high energy physics concerns the
possible existence of asymptotically important contributions to elastic
reactions coming from the C-odd Regge trajectory known as the
{\it Odderon} [1]. After the recent findings of the UA4/2
Collaboration [2] which finds $\bar \rho \equiv {{Re \, F_{\bar p p}(s, t=0)}
\over {Im \, F_{\bar p p}(s, t=0)}} = 0.135 \pm 0.015$ at $\sqrt s = 550 \,
GeV$, many authors [3, 4, 5] prefer to discuss the $pp$ and $\bar p p$ data
taking into account the {\it Pomeron} contribution only (and secondary
Reggeon contributions, of course). The problem of the C-odd amplitude, however,
remains entirely open both from the theoretical as well as from the
experimental point of view. Indeed, if the Pomeron is interpreted as the
trajectory interpolating C-even glueballs, unavoidably we are led to conjecture
that a trajectory interpolating C-odd glueballs should similarly exist; next,
pQCD
calculations in the semihard region $ m^2 \ll -t \ll s$ suggest an Odderon
intercept larger than unity as for the Pomeron [6]. In addition, it turns out
that an Odderon contribution is empirically absolutely necessary in order to
obtain a high quality fit to all existing high energy elastic $pp$ and
$\bar p p$ data [7]. Its contribution, essentially hidden by the dominating
Pomeron in the small $|t|$ region, becomes important at large $|t|$ in
${{d\sigma}\over{dt}}$.

The question which we would like to answer in the positive is whether
 qualitative differences are to be expected at $t=0$ if
the Odderon exists. In the following, we {\it assume} that a
C-odd contribution is asymptotically present; in this case, we will derive a
number of correlations between the various observables at $t=0$.
It is our contention that these correlations should already be able
to provide an indication of the Odderon provided we go to sufficiently high
energies. We believe that LHC energies will definitely show such correlations
and that even at RHIC energies we may already have an indication of them. This
makes a qualitative difference with respect to previous
considerations aiming at detecting the Odderon.

For simplicity, we begin with an explicit Ansatz and we later generalize our
considerations so as to make our conclusions model independent.

1.- Let us assume the following {\it specific} asymptotic behaviors for the
{\it Symmetric} (S) and {\it Antisymmetric} (A) amplitudes in which the $pp$
and $\bar p p$ amplitudes can be decomposed
$$F_S \equiv {1 \over 2} (F_{\bar p p} + F_{pp}) = i s \left [ A\, (ln \tilde
s)
\, ln \left ({{B \, {\tilde s}^{\Delta}} \over {ln \tilde s}} \right ) -
{{D_f}
\over { \sqrt {\tilde s}}} \right ],
\eqno(1)$$
$$F_A \equiv {1 \over 2} (F_{\bar p p} - F_{pp}) = - s \left [ \epsilon \,
ln^a \tilde s - {{D_{\omega} } \over {\sqrt {\tilde s}}}\right ],\eqno(2)$$
where $F \equiv F(s, t=0)$, $\tilde s = s \exp(-i \pi /2)$ and A ($\ge 0$),
B ($\ge 0$), $\Delta $ ($\ge 0$),
D's, $a \,$  and $\epsilon$ are some real coefficients ($0< a \le 1$
is within the Regge-eikonal model the maximal behavior of $F_A$ compatible with
s-channel unitarity [8])
and the subscripts $f$
and $\omega$ denote  the subasymptotic C-even and C-odd Reggeons. A dimensional
scale factor $s_0$ has been set =1 in all previous (and following) equations.
Then we have, for large $s$
$$ {1\over {8 \pi}}\left [\sigma_{tot}^{\bar p p}(s) - \sigma_{tot}^{pp}(s)
\right ]
\equiv {{\bar \sigma - \sigma} \over {8 \pi}} = {{Im F_A} \over s} \approx
\epsilon \, a \, {\pi \over 2} \, (ln \,s)^{a-1}
+ {D_{\omega} \over \sqrt{2s}}, \eqno(3)$$
$${{ Re F_A} \over s} \approx - \epsilon \, ln^a s + {D_{\omega}
\over \sqrt{2s}}.\eqno(4)$$
If $\Delta$ is small (say $\Delta \approx 0.1$ [9]), the Froissart-Martin
behavior [10, 11] for the total cross sections $\sigma$ and $\bar \sigma$ will
set in only at extremely high energies and will be $\propto A \Delta ln^2s$.
Given that the parameter $\epsilon$ (which according to our Ansatz {\it cannot
be identically zero} ) must anyhow be very small to comply with the data (say
$\epsilon \sim 10^{-2} - 10^{-3}$), up to energies of the order of the TeV
we will see the
decrease of ($\bar \sigma - \sigma$) due to the slow turning off of the
secondary Reggeons coupling. Only at $\sqrt s \ge 1 \, TeV$ can we expect it
to begin approaching the term $\propto \epsilon \, (ln \,s)^{(a-1)}$ predicted
by (3).
On the other hand, it is the term $\propto \epsilon \, ln^a s$ in (4) which is
needed [7, 1b, 1c] to account for the behavior of ${{d \sigma} \over {dt}}$
near
the dip region and at large $|t|$ values in order to have a qualitatively good
fit to the data. From our Ansatz (1, 2) we find $\bar \rho \sim {{A\, \Delta
\,\pi \,ln \, s- \epsilon \, ln^a s + {{D_{\omega} + D_f}\over {\sqrt {2s}}}
\over {{A\, \Delta \, ln^2s}}}}$, and $\rho \sim {{A\, \Delta \, \pi \, ln \, s
+ \epsilon \, ln^a s + {{D_f - D_{\omega}}\over {\sqrt {2s}}} \over {{A\,
\Delta
\, ln^2s}}}}$ (where $\bar \rho$ and $\rho$ are the ratios of the real to the
imaginary forward amplitudes for $\bar p p$ and $pp$ respectively). Hence,
at sufficiently high energies one predicts $\bar \rho < \rho$.

Most unfortunately, our knowledge of the $pp$ elastic data is presently
limited to the ISR energies $\sqrt s \approx 62 \, GeV$ which is much much
smaller than the scale $\approx 1 \, TeV$ where we expect these effects to
start showing up. At the ISR energies the values of $\rho$ and $\bar \rho$
are practically  the same (if anything, $\bar \rho \ge \rho$). It is,
however, most interesting that the high quality fit of Ref.[7] empirically
predicts $\bar \rho < \rho$ already at $\sqrt s \ge 100 \, GeV$ which is a
very good omen that at RHIC energies ($\sqrt s \approx 500 \, GeV$) we should
already see evidence of the asymptotic inequality $\bar \rho < \rho$.

2.- Let us now move to a more general case.

The full discussion of C-odd effects has been explicitly taken up in a simple
Regge-eikonal model with $\pom$ - $O$ weak degeneracy \footnote{$^1$}{$\pom$
for Pomeron and $O$ for Odderon.} [8, 12, 13], $\alpha_{\pom}(t)= \alpha_O(t)
 = 1 + \Delta + \alpha' t$ with unequal residues $\beta_{{\pom},O}
(t) = \lambda_{\pm} exp(\gamma_{\pm} t)$. This model reproduces (at least at a
qualitative level) all the prominent features of $F_{\bar p p}$ and $F_{pp}$
over the entire Regge domain ($s \gg 1 GeV^2$ and $0 \le -t \le const$) and
gives a prediction of new phenomena [13] which coincides with the predictions
coming from the extrapolations of the best $\chi^2$ fit to all high energy $pp$
and $\bar p p$ data [7]. In particular, at $t=0$, this model gives
a concrete realization of the amplitudes (1) and (2) with $a=1$. In effect,
the following exact expressions for $F_S$ and $F_A$ are obtained in the
limit $s \rightarrow \infty, \gamma_+
=\gamma_- \equiv \gamma$ (secondary Reggeon contributions omitted) [13]
$$F_S= is \left [ 2(\gamma + \alpha' \, ln \tilde s) ln\left ( {{e^C
\sqrt{\lambda_+^2 + \lambda_-^2} \, {\tilde s}^{\Delta}} \over {2( \gamma +
\alpha' \, ln \tilde s)}}\right ) \right],\eqno(5)$$
$$F_A = -s\left [2 \left (arctg {{\lambda_-} \over {\lambda_+}} \right)
\left (\gamma +\alpha' \, ln \tilde s \right) \right ]\eqno(6)$$
where $C \approx 0.577$ is the Euler constant. Thus, if the Pomeron and
Odderon intercepts are equal to $\Delta \approx 0.08$, and the ratio of the
Odderon/Pomeron couplings has the right order of magnitude
($\approx 10^{-2} - 10^{-3}$), this Regge-eikonal model gives a simple and
selfconsistent way to
take into account the C-odd effects in $pp$ and $\bar p p$ elastic collisions.
As already mentioned,in a Regge-eikonal model, the behavior $F_A \propto -s ln
\tilde s$ is the maximal one compatible with s-channel unitarity [8]. Let us
also note that for $s \gg 1$ the following correlation holds $(\bar \sigma -
\sigma) (\bar \rho - \rho) < 0$ which appears to be quite general and model
independent\footnote{$^2$}{If this relation is already known, we have not been
able to find it in the literature.}.

3.- The general case.

{}From analyticity (forward dispersion relations) and $s-u$ crossing symmetry,
we
know that it must be [14, 15] that
$$F_S=i s f(ln \tilde s), \quad \quad F_A = -s g(ln \tilde s)\eqno(7)$$
where $f(z)$ and $g(z)$ are some real functions of $z= ln \tilde s$ where [10,
11] $|f| \quad , |g| \le \, const \, ln^2 s$. We will assume that for
$s > s_1\gg 1$ the (otherwise arbitrary) functions $f(z)$ and $g(z)$ are
smooth (i.e.
not of an oscillatory character) functions belonging to the class of functions
that asymptotically (when $|z| \rightarrow \infty$) are $|z|^{-N} \le |f(z)|,
\,
|g(z)| \le const \, |z|^2$ where $N$ is any positive number. Then,
for $s > s_1$
 we can write
$$f(z) \approx f(x) \, - \, i {{\pi} \over 2} \, f'(x), \quad g(z) \approx g(x)
\, - \, i {{\pi} \over 2} \, g'(x),\eqno(8)$$
where $f(x),\, g(x), \, f'(x) = df(z)/dz|_{z=x}, \, g'(x) = dg(z)/dz|_{z=x}$
are real functions of $x=ln \, s$.

With the above definitions (and using the optical theorem), we have for the
observables introduced previously
$${{Re \, F_S} \over {Im \, F_S}} \, = \,
{{\bar \rho \, \bar \sigma \, + \, \rho \, \sigma} \over
{\bar \sigma \, + \, \sigma}} \approx {\pi \over 2} \,
{{f'} \over f}; \quad  {{Re \, F_A} \over {Im \, F_A}} \, = \,
{{\bar \rho \, \bar \sigma \, - \, \rho
\sigma} \over { \bar \sigma \, -\, \sigma}} \, \approx - {{2} \over {\pi}} {{g}
\over
{g'}};\eqno(9)$$
$$\bar \sigma \approx  4 \, \pi \, (f + {{\pi} \over 2} \, g') ;
\quad \quad \sigma \approx 4 \, \pi \, ( f - {{\pi} \over 2} \,
g');\eqno(10)$$
$$\bar {\rho} \approx {{{{\pi} \over 2} \, f' \, - \, g} \over
{f \, + \,  {{\pi} \over 2} \, g'}}; \quad \rho \approx
{{{{\pi} \over 2} \, f' \, + \, g} \over {f \, - \,
{{\pi} \over 2} \, g'}} \eqno(11)$$
where $f, g, f'$ and $g'$ stand for $f(x), g(x), f'(x)$ and $g'(x)$.

We can now define the following ratios of sums and differences of the
quantities given above
$$RD(\rho, \sigma) \equiv {{ ( \bar \rho - \rho )} \over
{ (\bar \sigma - \sigma )}} \, \approx \, -
{{ \bar \sigma + \sigma} \over { \pi  \, \bar \sigma \, \sigma}}
{g \over g'} \left [ 1 \, +\,  {{\pi}^2 \over 4} \,
{{f'} \over f } \, {{g'} \over g} \right ],\eqno (12)$$
$$RS(\rho, \sigma) \equiv {{\left ( \bar \rho + \rho \right )}\over
{  \left ( \bar \sigma + \sigma \right )}} \, \approx \,  {{ 2 {\pi}^2} \over
{ \bar \sigma \, \sigma}} \left [ f'\, +\,
{{g \, g'} \over f}  \right ].\eqno (13)$$
First of all, notice that, from (10) we clearly have
$f \,> \, {\pi \over 2} |g'|$; thus, while it is well known that $\rho$ will
remain (asymptotically) positive, $\bar \rho$ will be (asymptotically) positive
only if ${\pi \over 2} \, f' \, >  |g|.$
Next, notice that the ratios $f' \over f$ and $g' \over g$ decrease
always at least as $const / ln \, s$ (irrespective of whether $f$ and $g$ grow,
tend to a constant or decrease),
hence, the second term in square bracket in eq.(12)
can always be neglected asymptotically and the latter becomes
$$RD(\rho, \sigma) \approx   - {{ \bar \sigma + \sigma} \over
{ \pi  \, \bar \sigma \, \sigma}} {{g} \over {g'}} \approx -
{{ \bar \sigma + \sigma}  \over { \pi  \, \bar \sigma \, \sigma}}
{{ln \, s} \over K}, \eqno(14)$$
where $K$ is some constant.

Let us suppose first that $g$ grows to infinity (i.e. we have an increasing
Odderon contribution); in this case $ g' \over g $ is positive which implies
$K$ is positive. As a consequence, $RD(\rho, \sigma) < 0$. This is a direct
consequence of eqs.(9). As already mentioned, this is the case found in Ref.
[7] when extrapolating to higher energies and $|t|$ the fits to all existing
$pp$ and $\bar p p$ data . We expect that the change of sign of
 $RD(\rho, \sigma) < 0$ should already be observable at RHIC energies (if
the measurements will not be precise enough, however, one may have to wait for
LHC for an unambiguous answer). If, on the contrary, $g$ decreases to zero as
$s \rightarrow \infty$, then ${{g'} \over g}$ is negative, $K$ is thus negative
and $RD(\rho, \sigma) > 0$. The last possible option is when $g$ tends to a
constant in which case both ${{g'} \over g}$ positive or negative can
occur. In this case, however, another correlation exists, namely
$\left ( \bar \sigma' - \sigma' \right ) \, \left (\bar \rho - \rho \right )
> 0$ where we have defined $\sigma' \equiv {{d \sigma} \over {dt}}|_{t=0}=
{{1 \over {16 \, \pi}}} \, {\sigma}^2 \, \left (1+{\rho}^2 \right ).$

To see this in some detail, let us define
$$ 16 \, \pi \, {{\left ( \bar \sigma' - \sigma' \right ) } \over
{( {\bar \sigma}^2 - {\sigma}^2)}} \equiv RD(\sigma', \sigma) \equiv
\left [ 1 + {{ (\bar \sigma \, \bar \rho)^2 - (\sigma \, \rho)^2} \over
{{\bar \sigma}^2 - {\sigma}^2}} \right ] \approx \left ( 1 -  {f' \over f}
{g \over {g'}}\right ). \eqno(15)$$
When both $f$ and $g$ grow and
$f \over g$ also grows, we have  $ {f' \over f} \, {g \over g'} > 1$
and hence $RD(\sigma' , \sigma) \rightarrow -B$ where
$B$ is some positive coefficient. As an example, one can consider the case
$f(x)=A \, ln^2 s, \, \, g(x)= B \, ln^{\alpha} s, \, where \, 0< \alpha < 2$
(in this case, we have also $(\bar \sigma - \sigma)(\bar \rho - \rho) <0$).
When both $f$ and $g$ grow and ${f \over g} \rightarrow const$,
$RD(\sigma' , \sigma) \rightarrow 0$. As an example, consider the case
$f(x) = A \, ln^{\alpha} s$ and $g(x) = B \, ln^{\alpha} s$ where $0 < \alpha
\le 2$. Finally, when $f$ and $g$ both grow but  $f \over g$ decreases, then
we have $RD(\sigma' , \sigma) \rightarrow B$ where $B$ is some positive
number\footnote{$^3$}{This latter case is of no physical interest since it
would correspond to an Odderon asymptotically dominating over the Pomeron.}.
As an example, consider $f(x) = A \, ln^{\alpha} s$ and $g(x) =
B \, ln^{\beta} s$ with $\beta > \alpha$. Next, when $f$ grows and
$g \rightarrow const$, we have $RD(\sigma', \sigma) < 0$  if
${g' \over g} >0$ and $RD(\sigma', \sigma) > 2$ when ${g' \over g} < 0$.
Hence, $RD(\sigma', \sigma) \, RD(\rho, \sigma) > 0$ or
$(\bar {\sigma'} - \sigma') \, (\bar \rho - \rho)$ is positive.

When $f$ grows and $g$ decreases, we have $RD(\sigma', \sigma) \rightarrow
const > 1$. When $f \rightarrow const$ and $g$ grows, we have
$RD(\sigma', \sigma) >0$ (this case was first considered by Eden [16]). When
both $f$ and $g$ are growing, the quantity (13) $RS(\rho, \sigma) >0$. And
so on; all possibilities are easy to take into account and all correlations
follow quite straightforwardly.

Let us observe that the above correlations still persist when the leading
contributions cancel exactly between $pp$ and $\bar p p$ and we are left
purely with secondary Reggeons i.e. when $g(\tilde s) \approx const \,
(\tilde s)^{-{1 \over 2}}$ .

Summarizing, we have shown that a very sensitive indicator to a
large C-odd contribution in the amplitude is the sign of $RD(\rho, \sigma)$
while the correlation $RD(\sigma' , \sigma)$ tells us about the relative
value of the  C-odd contribution compared to the C-even one. Another
correlation which we have not been able to prove exactly but which we strongly
suspect to be correct, in the case of large asymptotic C-odd contribution is
$RD(B, \sigma) \equiv  {{\bar B - B} \over {\bar \sigma - \sigma}} > 0$ where
$B(s)$ is the slope of the diffraction peak defined as $B(s) \, = \,
{{d (ln {{d \sigma} \over {dt}}) }\over {dt}}|_{t=0}$. This relation holds true
in the Regge-eikonal model with $\pom $-$O$ degeneracy of Ref.[13].

In conclusion, somewhat to our surprise,
we have provided a host of correlations between $t=0$ quantities
(total cross
sections, ratios of real to imaginary forward amplitudes and forward angular
distributions) whose sign, in particular, appear quite sensitive to the
existence of a growing Odderon contribution. This analysis
goes quite beyond the conclusions reached in [7] where it was shown that
a C-odd contribution affects rather strongly the
{\it large}-$|t|$ data; our present results show, in fact, that
already $|t|=0$ data may be able to decide about the
existence of a relevant Odderon or not. Moreover, if the example of Ref.[7] can
 be taken as a guide, these correlations should already prove valid
at RHIC energies [17]. The ultimate test, however, should of course come from
LHC [18]. Experiments planned for these machines should check our correlations.

\medskip
\noindent{{\bf Aknowledgements}

\noindent{Two of us (M. G. and A. S.) would like to thank the Istituto
Nazionale di Fisica Nucleare (INFN) and
the Ministry of Universities, Research and Scientific Technology (MURST) of
Italy for their financial support while at the Department of Theoretical
Physics of the University of Torino where this work was done.}

\centerline{\bf References}

\vskip 10 pt
\item{[1]} a) L. Lukaszuk and B. Nicolescu, Nuovo Cimento Lett. {\bf 8} (1973)
405;

\item{} b) P. Gauron, E. Leader and B. Nicolescu, \PRL {\bf B54}{1985}{2656}
and \PRL{\bf B55}{1985}{639};

\item{} c) P. Gauron, B. Nicolescu and E. Leader, \NP {\bf B299}{1988}{640},
\PL
{\bf B238}{1990}{406}.

\smallskip
\item{[2]} C. Augier {\it et al.}, \PL {\bf B316}{1993}{448}.

\smallskip
\item{[3]} A. Martin,  {\it Do we understand total cross sections?}, Proc.
``Les Rencontres de Physique de la Vall\'ee d'Aoste", La Thuile, Aosta (Italy),
6-14 March 1994, p.407.
\smallskip
\item{[4]} J. Dias de Deus and A. Braz de Padua, \PL {\bf B317}{1993}{428}.
\smallskip
\item{[5]} E. Gottsman, E. M. Levin and U. Maor, \ZP{\bf C57}{1993}{677}.
\smallskip
\item{[6]} P. Gauron, L. Lipatov and B. Nicolescu, \PL {\bf B304}{1993}{334}.
\smallskip
\item{[7]} P. Desgrolard, M. Giffon and E. Predazzi \ZP{\bf C63}{1994}{241}.
\smallskip
\item{[8]} J. Finkelstein, H. M. Fried, K. Kang and C-I Tan, \PL {\bf B232}
{1989}{257};

\item{} E. S. Martynov, \PL {\bf B232}{1989}{367}.
\smallskip

\item{[9]} A. Donnachie and P. V. Landshoff, \NP {\bf B231}{1984}{189}, \NP
{\bf B244}{1984}{322}, \NP {\bf B267}{1986}{690}.
\smallskip
\item{[10]} M. Froissart, \PR {\bf 123}{1961}{1053}.
\smallskip
\item{[11]} A. Martin, \NC {\bf 42}{1966}{930}, \NC {\bf 42}{1966}{1219}.

\smallskip
\item{[12]} V. A. Petrov and A. P. Samokhin, \PL{\bf B237}{1990}{500}; Proc.
XXV Rencontres de Moriond (1990), p.411; Proc. XXVI Rencontres de Moriond
(1991), p.403, Ed. J. Tranh Than Van (Editions Fronti\`eres, Gif-sur-Yvette).
\smallskip
\item{[13]} A. P. Samokhin, Talk given at the XVIII International Workshop on
High Energy Physics and Field Theory, 26-30 June 1995, Protvino, Russia.
Proc. of the LP-HEP'91 Conference,Geneva 1991, vol.1,p.771, Ed. S.He-
garty,K.Potter and E.Quercigh (World Scientific).exit

\smallskip
\item{[14]} N. N. Meiman, Zh. Eksp. Teor. Fiz. {\bf 43}, 1962, 227 ( See also
translation in \JETP {\bf 16}{1963}{1609}).
\smallskip
\item{[15]} A. A. Logunov, Nguyen Van Hieu, I. T. Todorov and O. A. Khrustalev,
\PL {\bf 7}{1963}{69}.
\smallskip
\item{[16]} R. J. Eden, \PR {\bf D2}{1970}{529}.
\smallskip
\item{[17]} W. Guryn {\it et al.}, {\it Proposal to measure total and elastic
$pp$
cross sections at RHIC} presented at 6th Blois International Conference on
Elastic and Diffractive Scattering, Blois (France) 24-26 June 1995. To be
published by Editions Fronti\`eres.

\smallskip
\item{[18]} M. Buenerd {\it the TOTEM Collaboration}, {\it The TOTEM project at
LHC}
presented at 6th Blois International Conference on
Elastic and Diffractive Scattering, Blois (France) 24-26 June 1995. To be
published by Editions Fronti\`eres.

\end